\pdfoutput=1 % enforce pdflatex on arXiv /Fred

\documentclass[unnumsec,webpdf,contemporary,large,numbers,compress]{oup-authoring-template}

\usepackage[utf8]{inputenc}
\usepackage[T1]{fontenc}
\usepackage[english]{babel}

\usepackage{amsmath}
\usepackage{amssymb}

\usepackage{graphicx}
\usepackage[dvipsnames]{xcolor}
\usepackage{ulem} 
\usepackage{dsfont}
\usepackage[multidot]{grffile}

\usepackage{physics}

\newcommand{\mat}[1]{\boldsymbol{\mathrm{#1}}}

\newcommand{\ii}{\mathrm{i}}
\newcommand{\ee}{\mathrm{e}}
\newcommand{\kB}{k_\mathrm{B}}

\newcommand{\bS}{\begin{subequations}}
\newcommand{\eS}{\end{subequations}}

\newcommand{\lL}{{\mathsf L}}
\newcommand{\lR}{{\mathsf R}}

\newcommand{\da}{{\mathsf a}}
\newcommand{\db}{{\mathsf b}}

\newcommand{\vd}{{\vphantom{\dagger}}}

\newcommand{\LPT}{{LPT$_\Gamma$}} % für die Diskussion um fig.5

\DeclareMathOperator{\diag}{\mathrm{diag}}
\newcommand{\figWidth}{.9\linewidth}

\newcommand{\societylogo}{}
\begin{document}
\journaltitle{Journal}
\DOI{}
\copyrightyear{YEAR}
\pubyear{YEAR}
\vol{XX}
\issue{x}
\access{}
\appnotes{}

\firstpage{1}

\author[$\ast$]{Oliver Oing\ORCID{0009-0009-5795-4394}}
\author{Alexander Hahn\ORCID{0000-0002-9305-1768}}
\author{Jürgen König\ORCID{0000-0003-3836-4611}}
\author{Alfred Hucht\ORCID{0000-0002-9276-0159}}

\title[]{
    Aharonov-Bohm oscillations in transport through interacting quantum dots 
}

% \address{                    
%   %\inst{1} 
%   Theoretische Physik, Universität Duisburg-Essen and CENIDE, 47048 Duisburg, Germany
% }
\address{\orgdiv{Theoretische Physik}, \orgname{Universität Duisburg-Essen and CENIDE}, \orgaddress{\street{Lotharstr. 1}, \postcode{47048 Duisburg}, \country{Germany}}}
%\state{NRW}, 

\corresp[$\ast$]{Corresponding author: \href{email:oliver.oing@uni-due.de}{oliver.oing@uni-due.de}}

\received{Date}{0}{Year}
\revised{Date}{0}{Year}
\accepted{Date}{0}{Year}

\abstract{
We investigate electronic transport through a two-terminal Aharonov-Bohm interferometer with two quantum dots in the presence of finite intra-dot Coulomb interactions.
With the help of a diagrammatic technique, we derive, within linear perturbation theory in the tunnelling-coupling strength, analytic expressions for the occupations, current, and interferometric visibility in terms of a coherence measure that involves the sum of four off-diagonal elements of the reduced density matrix.
The latter are calculated numerically from a master equation for arbitrary settings of the system parameters.
For some symmetric setups, we obtain closed-form expressions for the linear conductance.
We systematically analyse the impact of a finite Coulomb interaction on the interferometric visibility, and reveal a rapid suppression of destructive interference upon breaking of system symmetries.
Our results are supported by alternative methods, including the TraSPI method.
}


\maketitle

\section{Introduction}\label{sec:Introduction}
Quantum coherence is a fundamental property of quantum mechanics with significant potential for a wide range of applications, including quantum materials \cite{lu_rational_2025,hwang_phase-controlled_2024} and qubits \cite{lo_enhancement_2025,swain_encapsulation_2025}, to name only a few.
One of the most prominent methods for probing quantum coherence is Aharonov-Bohm (AB) interferometry.
The AB effect has been studied in various contexts, including its interplay with Andreev bound states \cite{zalom_andreev_2026,pan_effects_2008}, quantum Hall systems \cite{kim_aharonovbohm_2026,biswas_anomalous_2024}, spin-dependent processes \cite{ibragimov_spin-dependent_2026} and Majorana bound states \cite{yang_phase_2026,jin_aharonovbohm_2024}, among others.

In this work, we investigate an AB interferometer containing two single-level quantum dots (QDs) in a ring geometry.
The discrete energy levels of QDs offer tunable resonances and, therefore, insights on the coherence and the role of the Coulomb interaction \cite{yacoby_coherence_1995,aikawa_observation_2004}.
Systems similar to the one studied here have been realised experimentally \cite{schuster_phase_1997, holleitner_coherent_2001, holleitner_probing_2002, sigrist_magnetic-field-dependent_2004, sigrist_coherent_2007, hatano_aharonov-bohm_2011, chiatti_electron_2014, prosko_flux-tunable_2024}.
Due to the interplay of nonequilibrium effects, Coulomb interaction, and finite temperature, precluding exact analytical treatment,
the theoretical description is challenging.

We address the regime of weak tunnel coupling and, therefore, choose linear perturbation theory in the tun\-nelling-coupling strength $\Gamma$ (\LPT).
Other numerical methods, such as renormalization-group schemes \cite{schoeller_real-time_2000, metzner_functional_2012, jakobs_nonequilibrium_2007, schoeller_perturbative_2009, anderson_poor_1970} or the transfer-matrix summation of path integrals method (TraSPI) \cite{mundinar_transfer-matrix_2022}, take all orders of the coupling strength into account but come with significantly higher computational costs.
The results calculated here with \LPT{} are almost instant and, in addition, enable us to obtain analytic expressions.
To verify the qualitative correctness, we compare with the aforementioned TraSPI method for a system with one interacting and one non-interacting dot, in which the TraSPI computation is feasible.

Since even in the weak-coupling regime we are left with many parameters, former papers made simplifying assumptions such as equal tunnelling strengths, equal energy levels, the linear-response regime regarding the bias voltage, and either zero \cite{konig_coherence_2001, konig_aharonov-bohm_2002, kubala_flux-dependent_2002} or infinite Coulomb interaction \cite{konig_coherence_2001, konig_aharonov-bohm_2002}.
Thereby, important physics is missed which can be described by the extended calculations done here.

To this end, we employ the Keldysh formalism and linear perturbation theory in the tunnelling strength $\Gamma$ \cite{konig_aharonov-bohm_2002}.
Furthermore, we have implemented routines in \textit{Wolfram Mathematica} \cite{MMA14} to be able to evaluate hundreds of diagrams for Green's functions and transition rates.
Analytically, this approach allows us to calculate the electron occupations of the quantum dots, the current, and further derived quantities in terms of eight system parameters. Numerically, these quantities can be computed for arbitrary settings of all eleven parameters.

\section{Model} \label{sec:Model}
\begin{figure}
    \centering
    \includegraphics[width=\linewidth]{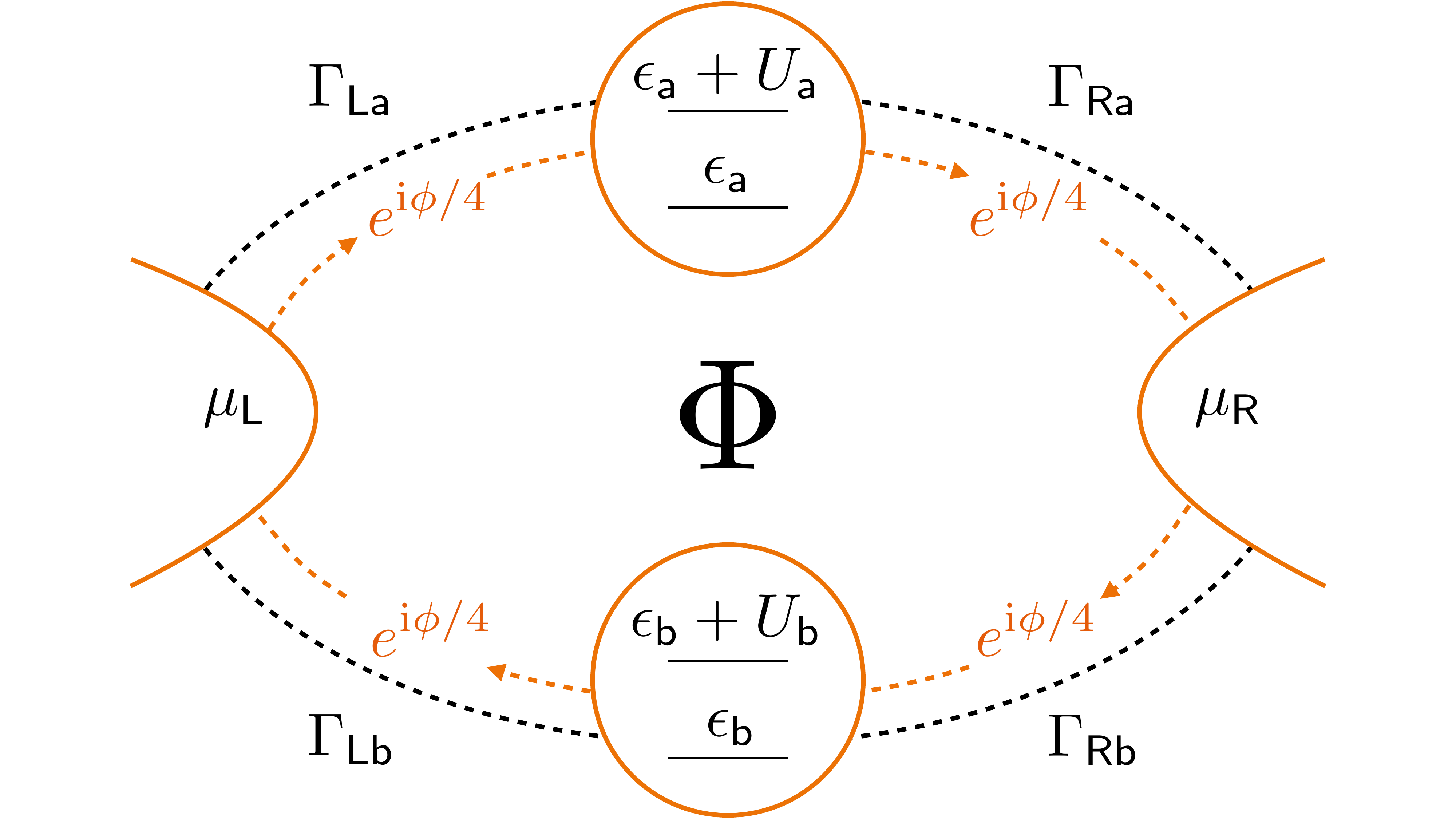}
    \caption{The two single-level quantum dots $q\in\{\da,\db\}$ in the AB interferometer couple to the two leads $\ell\in\{\lL,\lR\}$ with tunnelling strength $\Gamma_{\ell q}$.
    The enclosed magnetic flux $\Phi$ yields a phase $\phi$ in the tunnel couplings.
    }
    \label{fig:system}
\end{figure}
The considered AB interferometer contains two non-interacting leads and two interacting single-level QDs, as illustrated in fig.~\ref{fig:system}.
A bias voltage $V$ is applied between the two leads $\ell\in\{\lL,\lR\}$, represented through the chemical potential $\mu_\ell=-\ell eV/2$ where the lead index takes the values $\ell=\mp1$ for left and right lead, respectively, and $e$ is the electron charge.
Each QD $q\in\{\da,\db\}$ has a gate-tunable energy level $\epsilon_q$, measured from the Fermi edge of the leads in absence of the bias voltage $V$, and a Coulomb interaction $U_q$.

Electrons can tunnel between leads and QDs according to the respective tunnelling strengths $\Gamma_{\ell q} = 2\pi |t_{\ell q}|^2 N_\ell$, with tunnel matrix elements $t_{\ell q}$ and density of states $N_\ell$,
with negligible energy dependence in the wide-band limit.
Due to the magnetic flux $\Phi$ enclosed by the two interferometer arms, the tunnelling electrons acquire a phase $\phi=2\pi \Phi/\Phi_0$, with flux quantum $\Phi_0=h/e$, which can be distributed equally across all four tunnelling strengths by using the gauge freedom.

The AB interferometer is described by the Hamiltonian
\begin{align}\label{eq:Hamiltonian}
	\mathcal{H} &= \sum_{q \sigma} \epsilon^\vd_{q} \hat n^\vd_{q \sigma} + \sum_{q} U^\vd_q \hat n^\vd_{q \uparrow} \hat n^\vd_{q \downarrow}
    + \sum_{\ell \vb{k} \sigma} \epsilon^\vd_{\ell \vb{k}} \hat a^\dagger_{\ell \vb{k} \sigma} \hat a^\vd_{\ell \vb{k} \sigma}\nonumber\\
	&\quad + \sum_{ \ell \vb{k} q \sigma} \big(t^\vd_{\ell q} \hat a^\dagger_{\ell \vb{k} \sigma} \hat c^\vd_{q  \sigma} + \text{h.c.}\big) \, ,
\end{align}
where $\hat n^\vd_{q \sigma} = \hat c^{\dagger}_{q \sigma} \hat c^\vd_{q \sigma}$ is the occupation number operator for electrons with spin $\sigma\in\{\uparrow,\downarrow\}$ on dot $q$, with the creation and annihilation operators $\hat c^{\dagger}_{q \sigma}$ and $\hat c^\vd_{q \sigma}$.
The electrons in lead $\ell$ with energy $\epsilon_{\ell \vb k}$ and momentum $\vb k$ are described by the corresponding operators $\hat a^\vd_{\ell \vb{k} \sigma}$, while the last term describes the tunnel coupling. 
The two QDs are assumed to be sufficiently separated, such that direct (tunnelling or electrostatic) interactions are irrelevant and, therefore, are not taken into account.

We introduce $\chi=\alpha\beta$ to denote the state of the system, which combines the states $\alpha, \beta$ for the two dots $\da,\db$, respectively.
Each dot can be empty ($\alpha,\beta=0$), singly occupied with spin $\sigma$ ($\alpha,\beta=\sigma$), singly occupied with the opposite spin $\bar\sigma$ ($\alpha,\beta=\bar \sigma$), or doubly occupied ($\alpha,\beta=\text{d}$).
In total, this yields 16 possible system states $\chi$.

\section{Method} \label{sec:Method}
For non-interacting leads, the current from the dots to one lead $\ell$ in terms of Green's functions can be expressed as \cite{meir_landauer_1992,konig_aharonov-bohm_2002}
\begin{align}\label{eq:currentG}
    I_\ell &= \frac{e}{\ii h} \tr 
    \mat{\Gamma}_\ell \sum_\sigma \int \mathrm d \omega \Big(\bar f_\ell \mat G_\sigma^{<} + f_\ell \mat G_\sigma^{>} \Big) \, .
\end{align}
The occupation of the leads is given by the Fermi distribution $f_\ell(\omega)=1/(\exp[(\omega-\mu_\ell)/\kB T]+1)$, and we write its complement as $\bar f_\ell(\omega)=1-f_\ell(\omega)$.

The matrix product of the tunnelling-strength matrix $\mat{\Gamma}_\ell$ and the lesser and greater Green's functions matrix $\mat G_\sigma^\lessgtr$ yield all possible combinations of tunnelling.
Defining the geometric mean $\tilde\Gamma_\ell=\sqrt{\Gamma_{\ell \da} \Gamma_{\ell \db}}$, the matrices
\begin{align}\label{eq:matGamma}
    \mat\Gamma_\ell=\begin{pmatrix}
        \Gamma_{\ell \da} & 
        \tilde\Gamma_\ell \, \ee^{-\ii \ell \phi/2} \\
        \tilde\Gamma_\ell \, \ee^{\ii \ell \phi/2} & \Gamma_{\ell
         \db}
    \end{pmatrix}
\end{align}
contain all tunnel couplings and the magnetic flux-depend\-ent phase $\phi$.
On the other hand, the lesser and greater Green's function matrices contain all possible combinations of the two dots,
\begin{align}\label{eq:matG}
    \mat{G}_\sigma^\lessgtr(\omega) = 
    \begin{pmatrix}
        G_{\vphantom{\db}\da \da \sigma}^\lessgtr(\omega) & 
        G_{\da \db \sigma}^\lessgtr(\omega) \\[3pt]
        G_{\db \da \sigma}^\lessgtr(\omega) & 
        G_{\db \db \sigma}^\lessgtr(\omega)
    \end{pmatrix}.
\end{align}
The Fourier-transformed elements of \eqref{eq:matG} are given by the expectation values
\bS\begin{align}\label{eq:Glegr}
    \tilde G_{q q' \sigma}^{<}(t)&=+i\big\langle c^\dagger_{q' \sigma}(0) \, c^{\vphantom{\dagger}}_{q \sigma}(t)\big\rangle \, ,\\
    \tilde G_{q q' \sigma}^{>}(t)&=-i\big\langle c^{\vphantom{\dagger}}_{q \sigma}(t) \, c^\dagger_{q' \sigma}(0)\big\rangle \, .
\end{align}\eS
Combining the currents for both leads, we finally obtain the total current $I=(I_\lR-I_\lL)/2$ through the system.
\subsection{First-order transport}
We choose the same approach as done in ref. \cite{konig_aharonov-bohm_2002} and expand \eqref{eq:currentG} to first order in the tunnelling strength $\Gamma$, neglecting terms $\mathcal O(\Gamma^2)$.
Thus, the calculations describe single-electron tunnelling processes, while cotunnelling processes are not taken into account.
The Green's functions appear solely in zeroth order, which we compute using a diagrammatic approach (for details see \cite{konig_aharonov-bohm_2002}).
These Green's functions $G^\lessgtr_{q q' \sigma}$ are connected to density matrix elements $\rho^\chi_{\chi'}=\langle\, \ketbra{\chi'}{\chi}\,\rangle$, which are subsequently solved via master equations as described below.

We start with the Green's functions on the matrix  diagonal in \eqref{eq:matG}, containing only diagonal density matrix elements $\rho ^\chi_\chi$ which represent the probability of the system being in the state $\chi$,
\bS\label{eqs:Gd}\begin{align}\label{eq:Gdlesser}
    G^{<}_{q q \sigma} &= + \ii \pi \big[
         P_{q 1}\delta(\omega-\epsilon_q)+
        2P_{q 2}\delta(\omega-\epsilon_q-U_q)\big] \, ,\\
    \label{eq:Gdgreater}
    G^{>}_{q q \sigma} &= - \ii \pi \big[
        2P_{q 0}\delta(\omega-\epsilon_q) + 
         P_{q 1}\delta(\omega-\epsilon_q-U_q)\big] \, .
\end{align}\eS%
Here, $\delta(\cdot)$ is the Dirac delta distribution, and $P_{q n}$ represents a combination of diagonal density matrix elements where dot $q$ is occupied with $n$ electrons.
For example, $P_{\da 0}=\sum_\beta \rho^{0\beta}_{0\beta}$ contains all diagonal density matrix elements where the system is in a state with unoccupied dot $\da$ ($\alpha=0$).
Summing up the probabilities for all three charge occupations equals one, thereby satisfying the normalization condition.
Furthermore, the mean dot occupation is given by 
\begin{align}
    N_q = \sum_n n \, P_{q n} = P_{q 1}+2P_{q 2} \, .
\end{align}%

From the definitions \eqref{eqs:Gd} as well as \eqref{eq:currentG}, it becomes clear that the lesser Green's functions are associated with electron tunnelling from the QD to the leads, while the  greater Green's functions describe the opposite processes. 
The off-diagonal Green's functions $G_{q \bar q \sigma}^\lessgtr$ in \eqref{eq:matG} couple to off-diag\-onal density matrix elements, or \textit{coherences}, $\rho ^\chi_{\chi'}$.
To zeroth order, lesser and greater Green's functions are equal, such that the labels $\lessgtr$ can be omitted,
\begin{align}\label{eq:Gab}
    G_{q \bar q \sigma} = G^<_{q \bar q \sigma} = G^>_{q \bar q \sigma} \, .
\end{align}
They are explicitly given by
\begin{align}\label{eq:Goffdia}
    G_{\da \db \sigma} = \ii \pi \Big( 
        & \rho_{0\sigma}^{\sigma0}
            \big[\delta(\omega - \epsilon_\da) + \delta(\omega - \epsilon_\db)\big] \nonumber\\
    {}+{}&\rho_{\text{0d}}^{\sigma\bar\sigma} 
            \big[\delta(\omega - \epsilon_\da) + \delta(\omega - \epsilon_\db - U_\db)\big] \nonumber\\
    {}-{}&\rho_{\bar\sigma\sigma}^{\text{d0}}
            \big[\delta(\omega - \epsilon_\da - U_\da) + \delta(\omega - \epsilon_\db)\big] \nonumber\\
    {}-{}&\rho_{\bar\sigma\text{d}}^{\text{d}\bar\sigma}
            \big[\delta(\omega - \epsilon_\da - U_\da) + \delta(\omega - \epsilon_\db - U_\db)\big]\Big) \, ,
\end{align}
while $G_{\db \da \sigma}=-G_{\da \db \sigma}^\ast$ due to the Hermiticity $\rho^{\chi'}_{\chi} = \rho^{\chi\,\ast}_{\chi'}$ of the density matrix.

For convenience, we introduce the \textit{effective coherence}
\bS\label{eqs:Nab}\begin{align}
    C_{\da\db}(\phi)&=\frac{1}{2\pi\ii} \sum_\sigma \int \dd\omega \, G_{\da \db \sigma}\\
    \label{eq:Nab2}
    &=\sum_\sigma\big(\rho ^{\sigma 0}_{0 \sigma}+\rho ^{\sigma \bar\sigma}_{0\text{d}} - \rho ^\text{d0}_{\bar\sigma \sigma} - \rho ^{\text{d}\bar\sigma}_{\bar\sigma \text{d}} \big) \, .
\end{align}\eS%
Other off-diagonal density matrix elements such as $\rho ^\text{d0}_\text{0d}$ do not contribute since the zeroth-order Green's functions contain only one creation and one annihilation operator.
Therefore, the coherences in \eqref{eq:Nab2} are all contributions necessary to describe the current and occupation in \LPT, thus making $C_{\da\db}(\phi)$ an effective coherence.

Consequently, using \eqref{eq:Gab} and performing the trace in the expression \eqref{eq:currentG} for the current, we can explicitly separate the contributions originating from the diagonal and from the off-diagonal Green's functions to get
\begin{align}\label{eq:currentlpt}
    I_\ell(\phi)
    &=\frac{2\pi e}{h} 
    \sum_{q} \left[ \big(N_q(\phi)-1\big)A^+_{\ell q} - A^-_{\ell q} + \Theta_\ell(\phi)
    \right].
\end{align}
Here, we introduced the phase-independent constants
\begin{align}\label{eq:Alq}
    A^\pm_{\ell q} &= \Gamma_{\ell q}\left[f_\ell(\epsilon_q) \pm \bar{f}_\ell(\epsilon_q + U_q)\right],
\end{align}%
which can be computed directly from the system parameters, as well as the explicitly phase-dependent \textit{interference measure}
\bS\label{eqs:Th}\begin{align}
    \Theta_\ell(\phi) &= \frac{\tilde\Gamma_\ell}{4\pi\ii} \sum_\sigma \int \dd\omega \, \Big( \ee^{\frac\ii2 \ell \phi} G_{\da \db \sigma} + \ee^{-\frac\ii2 \ell \phi} G_{\db \da \sigma} \Big)\\
    &= \tilde\Gamma_\ell \Re\Big[ C_{\da\db}(\phi) \, \ee^{\frac\ii2 \ell \phi} \Big].
\end{align}\eS%
Since all density matrix elements are implicitly phase dependent, the probabilities $P_{q n}$ and thus the occupation numbers are also contributing a phase dependency.
However, in the following section we show that we can express the full phase dependence of both the current $I(\phi)$ and the dot occupation numbers $N_q(\phi)$ through the interference measure $\Theta_\ell(\phi)$ from \eqref{eqs:Th}.

\subsection{Quantum master equations}\label{sec:qme}
To determine the density matrix elements, we use quantum master equations within a diagrammatic approach \cite{konig_aharonov-bohm_2002}:
we start with the stationary master equation in Liouville space
\begin{align}\label{eq:qme}
- \ii \big( \epsilon_{\chi} - \epsilon_{\chi'} \big) \rho ^{\chi}_{\chi'} + \sum_{\tilde\chi  \tilde\chi'} W^{\chi  \tilde\chi}_{\chi'  \tilde\chi'} 
\rho ^{\tilde\chi}_{\tilde\chi'}&=0 \, ,
\end{align}
where $\epsilon_{\chi^{\vphantom{\prime}}}$ denotes the energy corresponding to the state $\chi^{\vphantom{\prime}}$, and $W^{\chi  \tilde\chi}_{\chi'  \tilde\chi'}$ is the transition rate from the density matrix element $\rho ^{\tilde\chi}_{\tilde\chi'}$ to $\rho ^{\chi}_{\chi'}$.

The 16 possible states $\chi$ of our system result in 16 diagonal density matrix elements.
For the off-diagonal elements $\rho^\chi_{\chi^\prime}$, only spin- and particle-preserving elements must be taken into account. 
Therefore, the particle number and overall spin of $\chi$ and $\chi^\prime$ have to be equal.
This results in 20 remaining off-diagonal density matrix elements.
The spin symmetry further reduces the number of density matrix elements to 10 diagonal and 11 off-diagonal ones, resulting in a total of 21 elements.

The remaining density matrix elements are combined in a vector $\vec{\rho}$.
By applying eq.~\eqref{eq:qme} to each element, we can build the matrix equation
\begin{align}\label{eq:matEq}
    \big(\mat W - \ii\,\mat\diag\Delta\vec\epsilon \,\big) \vec{\rho} = \vec{0} \, ,
\end{align}
with the energy differences $\Delta \epsilon^\chi_{\chi'}=  \epsilon_\chi - \epsilon_{\chi'}$.
In addition, since the diagonal elements $\rho^\chi_\chi$ represent probabilities, they have to fulfil the normalization condition $\sum_{\chi} \rho ^\chi_\chi=1$.

In order to calculate the density matrix elements $\vec{\rho}$ with these equations, we need to determine the transition rates $\mat W$, which is done by a diagrammatic approach \cite{konig_aharonov-bohm_2002}.
Given that we are including eleven free parameters, hundreds of diagrams need to be evaluated.
Therefore, a \textit{Mathematica} routine was implemented as part of this work, which creates and evaluates all required diagrams.

By combining the rows of the linear equations corresponding to the combined density matrix elements $P_{q n}$, we can extract the expression for the phase-dependent occupation numbers of the two dots as
\begin{align}\label{eq:N}
    N_q(\phi) = 1 - \sum_{\ell} \frac{\Theta_\ell(\phi) - A^-_{\ell q}}{A^+_{\lL q} + A^+_{\lR q}} \, .
\end{align}%
The phase dependency of \eqref{eq:N} is solely given by the interference measure $\Theta_\ell(\phi)$ from \eqref{eqs:Th}, while all other contributions are described by the constants \eqref{eq:Alq}.
The only remaining unknown is the complex quantity $C_{\da\db}(\phi)$ entering $\Theta_\ell(\phi)$, for which the linear equations \eqref{eq:matEq} have to be solved.

Using \eqref{eq:Alq}, \eqref{eqs:Th}, and \eqref{eq:N}, the total phase-dependent current can be written as the compact expression
\begin{align}\label{eq:current}
    I(\phi) &= \frac{2\pi e}{h} \sum_{\ell q} \ell \, A^+_{\bar\ell q} \, 
    \frac{\Theta_\ell(\phi) - A^-_{\ell q}}{A^+_{\lL q} + A^+_{\lR q}} \, ,
\end{align}%
with $\bar\ell=-\ell$.
The phase-independent part is the naive sum of the currents through each dot, as if the two arms would not interfere.
Therefore, the destructive interference of the system is completely described by $\Theta_\ell(\phi)$.
Notably, the same fraction appears in both the current~\eqref{eq:current} and the occupation number~\eqref{eq:N}.
\section{Results} \label{sec:Results}

\begin{figure}
    \centering
    \includegraphics[width=\figWidth]{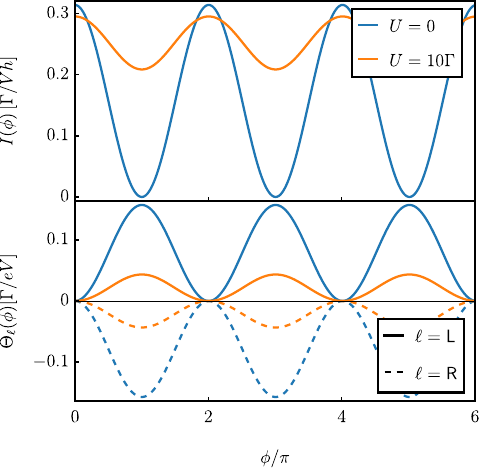}
    \caption{AB oscillations in the current $I(\phi)$ and in the interference measure $\Theta_\ell(\phi)$ as functions of the AB phase $\phi$, for a symmetric system with $\epsilon_q=-U_q/2=-U/2$, $U\in \{0,10\Gamma\}$, $\Gamma_{\ell q}=\Gamma$, $eV=10^{-3}\Gamma$ and $\kB T=10\Gamma$.}
    \label{fig:OverPhi}
\end{figure}%

The expressions \eqref{eq:N} and \eqref{eq:current} for the dots' occupation and the current define a main result of this work.
It is quite remarkable that, for \textit{any} choice of system parameters, both $N_q(\phi)$ and $I(\phi)$ can be expressed in such a simple form, where the interference measure $\Theta_\ell(\phi)$ is the \textit{only} ingredient which depends on density matrix elements.
Although only a special combination of these density matrix elements is needed, see eq.~\eqref{eq:Nab2}, one nevertheless has to solve the full $21\times 21$ system \eqref{eq:matEq}.
An analytic solution that takes all eleven parameters into account is possible but extremely time consuming and yields highly elaborated
expressions that offer no advantage over numerical solutions.

Compact analytic solutions are only available in special cases of high symmetry.
These include the linear conductance $G^\mathrm{lin}(\phi)=\dd I(\phi)/\dd V|_{V=0}$ in the case of identical tunnel couplings, $\Gamma_{\ell q}=\Gamma$, together with equal orbital energies, $\epsilon_q=\epsilon$, and either zero, $U=U_q=0$, or infinite, $U=U_q=\infty$, charging energy, that was studied earlier \cite{konig_aharonov-bohm_2002,konig_coherence_2001}.
We reproduce (after correcting a factor of 2 for a missing spin summation)
\begin{align}\label{eq:Glin}
    G^\mathrm{lin}_{0,\infty}(\phi)
    = -\frac{4 \pi e^2\Gamma}{h} f'(\epsilon) \left( B_{0,\infty}- B_{0,\infty}^3\sin ^2 \frac{\phi}{2} \right),
\end{align}
with $B_0=1$ for $U=0$ and $B_\infty=1/[1+f(\epsilon)]$ for $U=\infty$.
Here, $f(\epsilon)$ is the equilibrium Fermi function.

We now can extend this result to the case of a finite charging energy at the particle-hole symmetric point $\epsilon+U=-\epsilon$, and find the somewhat more involved expression
\begin{align}\label{eq:IAnalytic}
    G^\mathrm{lin}_\mathrm{symm}(\phi)
    &= -\frac{4 \pi e^2 \Gamma}{h} \left[ f'(\epsilon) \cos ^2 \frac{\phi}{2} \right. \nonumber \\ 
    &\left. \quad {}+ \frac{\epsilon^2 f'(\epsilon)- \epsilon\psi (\epsilon) }{\frac43 \epsilon^2 + 3 + \sin^2 \frac{\phi}{2} }\sin^2 \frac{\phi}{2} \right]  ,
\end{align}
where $\psi(\epsilon)=\Re\big[\frac{\partial}{\partial \epsilon}\Psi\big(\frac{1}{2}+\ii \frac{\epsilon}{2\pi \kB T}\big)\big]/\pi$ is an energy level renormalization \cite{splettstoesser_tunneling-induced_2012}, containing the digamma function $\Psi$.
The second term in the conductance~\eqref{eq:IAnalytic} vanishes for $\phi=0$, while for a phase of $\phi=\pi$ the first term becomes zero.

Beyond these special cases, we determine the current numerically.
Figure~\ref{fig:OverPhi} shows the influence of the Coulomb interaction on AB oscillations.
The blue line represents a system without Coulomb interaction and the orange line shows the particle-hole-symmetric case described in equation~\eqref{eq:IAnalytic} with energies $U=-2\epsilon=10\Gamma$.
In the current $I$, amplitude and average of the oscillations become smaller for finite Coulomb interactions and have a period of $2\pi$.
The current reaches its maximum at $\phi=0$ and its minimum at $\phi=\pi$, making these two phases the focus of subsequent analysis.
In both depicted cases, the destructive interference is fully described by the difference $\Theta_\lR(\phi)-\Theta_\lL(\phi)$, and the left and right contributions fulfil the simple relation $\Theta_\lL(\phi)=-\Theta_\lR(\phi)$.

However, for asymmetric configurations (for example $\epsilon_\da\neq\epsilon_\db$) this does not hold any more. In these cases, the sum over $\Theta_\ell(\phi)$ in the occupation number \eqref{eq:N} yields a finite contribution, making $N_q$ phase dependent.

Beyond the linear-response regime, the conductance $G(\phi)$ is calculated numerically as the slope between $I(\phi)$ at $V-\Delta V/2$ and $V+\Delta V/2$, and we used $e\Delta V=10^{-6}\Gamma$ in the following.
To visualise the influence of the AB phase, we introduce the \textit{interferometric visibility}, defined as the ratio 
\begin{align}\label{eq:Visibility}
    \nu=\frac{G(\phi=0)-G(\phi=\pi)}{G(\phi=0)+G(\phi=\pi)} \, .
\end{align}
It ranges from $\nu=+1$ (complete destructive interference at $\phi=\pi$) through $\nu=0$ (no interference at all) to $\nu=-1$ (complete destructive interference at $\phi=0$).

For the particle-hole symmetric case, eq.~\eqref{eq:IAnalytic} yields the visibility
\begin{align}\label{eq:AnalyticVisibility}
    \nu^\mathrm{lin}_\mathrm{symm} = \frac{\left(4+\frac13
   \epsilon ^2\right) f'(\epsilon) + \epsilon \psi(\epsilon)}{\left(4+\frac73
   \epsilon ^2\right) f'(\epsilon)- \epsilon \psi (\epsilon)} \, ,
\end{align}
which reaches its maximum $\nu=1$ at $\epsilon=0$ and monotonically decreases with increasing $|\epsilon|$.

\begin{figure}
    \centering
    \includegraphics[width=\figWidth]{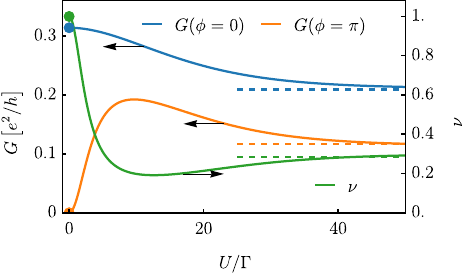}
    \caption{Behaviour of the conductance $G$ and the visibility $\nu$ for a symmetric system with finite Coulomb interactions $U$, $\epsilon=0$, $\Gamma_{\ell q}=\Gamma$, $eV=10^{-3}\Gamma$ and $\kB T=10\Gamma$. 
    The filled dots at $U=0$ and the dashed line (representing $U=\infty$) correspond to the analytic results from ref. \cite{konig_aharonov-bohm_2002}, the solid lines represent the numeric calculations done in this work.}
    \label{fig:UCrossover}
\end{figure}%
The calculations in ref. \cite{konig_aharonov-bohm_2002} were limited to the two special cases of either vanishing ($U=0$) or infinite ($U=\infty$) charging energy.
With our theory, we can now explore the full regime between these two limits.
This is demonstrated in fig.~\ref{fig:UCrossover}, which shows the conductances at $\phi=0$ and $\phi=\pi$ as well as the visibility as functions of $U_q=U$ for $\Gamma_{\ell q}=\Gamma$, $\epsilon_q=0$, $\kB T=10\Gamma$, and $eV=10^{-3}\Gamma$.
For a vanishing AB phase $\phi=0$, the conductance monotonically decreases with increasing $U$, simply because the charging energy suppresses transport channels that involve double occupancy of the dots.

At $\phi=\pi$, the variation of the linear conductance as function of $U$ is more pronounced.
In the non-interacting case $U=0$, perfect destructive interference completely suppresses transport.
For finite $U$, the conductance is substantially increased, since the charging energy renders the transport partially incoherent, such that destructive interference is less effective.
The subsequent decrease of the conductance, on the other hand, is once more due to an overall suppression of transport.
While the rapid increase is dominated by the tunnelling strength $\Gamma$, the decrease is thermally driven.

This leads to a non-monotonic dependence of the visibility $\nu$ on $U$, see fig.~\ref{fig:UCrossover}.
The numerical results agree with the analytic expressions $\nu=1$ for $U=0$ and approach those ($\nu=\frac27$) for $U=\infty$.

\begin{figure*}
    \centering
    % \begin{subfigure}[t]{0.33\linewidth}
        %\vskip 0pt
        %\frame{
        \includegraphics[height=195pt]{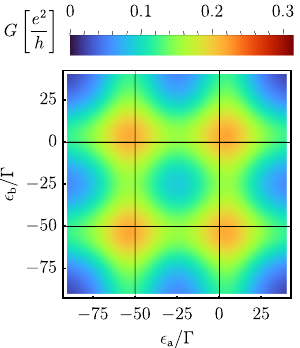}%}
        %\caption{Conductance $G$ for $\phi=0$\label{fig:G2DPhi0}}
    % \end{subfigure}\hfill%
    % \begin{subfigure}[t]{0.33\linewidth}
        %\vskip 0pt
        %\frame{
        \hspace{2pt}
        \includegraphics[height=195pt]{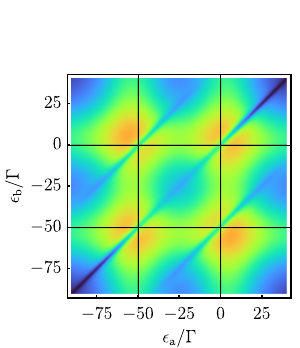}%}
        %\caption{Conductance $G$ for $\phi=\pi$\label{fig:G2DPhiPi}}
    % \end{subfigure}\hfill%
    % \begin{subfigure}[t]{0.33\linewidth}
        %\vskip 0pt
        %\frame{
        \hspace{2pt}
        \includegraphics[height=195pt]{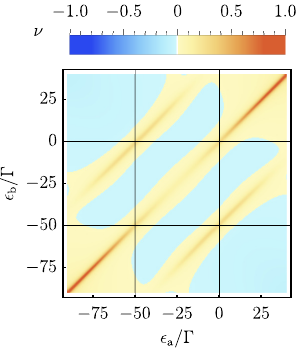}%}
        %\caption{Interferometric visibility $\nu$\label{fig:Visibility}}
    % \end{subfigure}
    \caption{\label{fig:U50energies}
    Conductance $G(\phi=0)$ (left), $G(\phi=\pi)$ (centre), and interferometric visibility (right) as functions of the energies $\epsilon_\da,\epsilon_\db$ for an AB system with symmetric Coulomb interactions $U=50 \Gamma$, symmetric tunnel couplings $\Gamma_{\ell q}=\Gamma$, a bias voltage of $eV=10^{-3}\Gamma$ and a thermal energy of $\kB T=10\Gamma$. 
    The black lines mark the resonances of QD $\da$ (vertical lines) and QD $\db$ (horizontal lines) for particle-like ($\epsilon_q=0$) and hole like ($\epsilon_q=-U_q$) excitation.}
\end{figure*}
The interpolation between the limits of $U=0$ and $U=\infty$ is only one virtue of our theory.
Very relevant from the experimental point of view is the possibility to tune the energies $\epsilon_\da$ and $\epsilon_\db$ \textit{independently} from each other, by which the symmetry between the two dots can be broken.
In fig.~\ref{fig:U50energies} (left), we show the conductance at $\phi=0$ for $U_\da=U_\db=U=50\Gamma$, $\Gamma_{\ell q}=\Gamma$, $eV=10^{-3}\Gamma$ and $\kB T=10\Gamma$ as a function of $\epsilon_\da$ and $\epsilon_\db$. 
In the limits $U=0$ and $U=\infty$, only \textit{one} resonance peak is expected at $\epsilon_\da=\epsilon_\db=0$ and $\epsilon_\da=\epsilon_\db=(2 \ln 2) \kB T$, respectively.
A finite value of $U$ changes this picture qualitatively.
Each dot has now \textit{two} excitation energies, that are associated with transport processes that we refer to as \textit{particle-like} at $\epsilon_q$ and \textit{hole-like} at $\epsilon_q+U_q$.
This results in \textit{four} peaks that lie close to the intersection points of the black lines that mark the resonances of either $\epsilon_q$ or $\epsilon_q+U_q$ with the Fermi edge of the leads.
At the peak on the top right, particle-like processes are involved in both dots, while at the bottom left, hole-like processes are the relevant excitations in both dots.
In addition, there are two peaks at the off-diagonal (top left and bottom right), in which one dot experiences particle-like and the other dot hole-like excitations.

While the conductance at $\phi=0$ mainly reflects the energy dependence of the transmission through the individual dots, the conductance at $\phi=\pi$, as shown in fig.~\ref{fig:U50energies} (centre), is suitable to probe the coherence.
Destructive interference leads to pronounced valleys in the conductance along $\epsilon_q=\epsilon_{\bar q}$ as well as $\epsilon_q+U_q=\epsilon_{\bar q}$, i.e., at the degeneracies of the excitation energies of the two dots.
For finite detuning, the destructive interference becomes suppressed and the conductance rapidly goes up.

We note that while the width of the resonance peaks is determined by the temperature $\kB T$, the width of the valley is governed by the tunnel coupling $\Gamma$ and, therefore, is much smaller in the presented example.
This is a direct consequence of the competition of $\Delta \epsilon$ and $\Gamma$ in the master equation \eqref{eq:matEq}.
As a result, the visibility, shown in fig.~\ref{fig:U50energies} (right), is strongly enhanced along the diagonals $\epsilon_q=\epsilon_{\bar q}$ and $\epsilon_q+U_q=\epsilon_{\bar q}$, while it remains small or even becomes negative in between.
This rich structure is completely overlooked when assuming fully symmetric quantum-dot setups as done in ref. \cite{konig_aharonov-bohm_2002}

The pronounced ridges in the visibility, together with a sign change of $\nu$ for larger detuning, is \textit{not} an artifact of the employed linear perturbation theory.
In contrast, the same behaviour is seen in the exact solution for the $U_q=0$ case \cite{kubala_flux-dependent_2002}, which includes processes of \textit{all} order in the tunnelling-coupling strength. 
Recalculating fig.~\ref{fig:U50energies} (right) with \LPT{} for $U=0$ yields qualitatively the same result as the exact solution.

\begin{figure*}
    \centering
    % \begin{subfigure}{0.33\linewidth}
    % height=168pt equals Fig.4
        %\frame{
        \includegraphics[height=160pt]{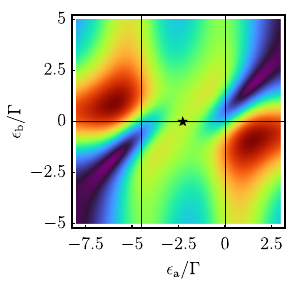}%}
        %\caption{Conductance $G$ (\LPT)}
        %\label{fig:TraSPIComp}
    % \end{subfigure}%
    % \begin{subfigure}{0.43\linewidth}
        %\frame{
        \hspace{2pt}
        \includegraphics[height=160pt]{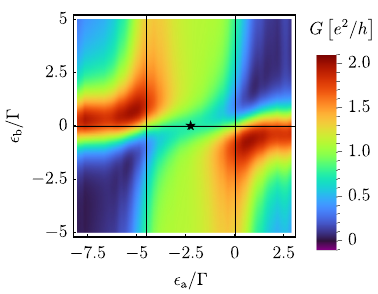}%}
        %\caption{Conductance $G$ (TraSPI)}
        %\label{fig:TraSPI}
    % \end{subfigure}%
    % \begin{subfigure}{0.24\linewidth}
        % \centering
        %\frame{
        \hspace{4pt}
        \includegraphics[height=160pt]{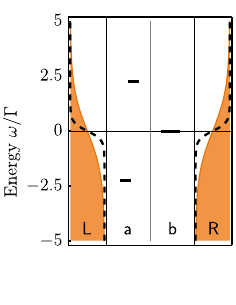}%}
        %\caption{Energy scheme}
        %\label{fig:energyPlot}
    % \end{subfigure}
    \caption{\label{fig:TraSPI}
    Conductance $G(\phi=\pi)$ as a function of the energies $\epsilon_\da,\epsilon_\db$, with charging energies $U_\da=4.5\Gamma$ and $U_\db=0$, symmetric tunnel couplings $\Gamma_{\ell q}=\Gamma$, and bias voltage $eV=10^{-2}\Gamma$.
    The thermal energy is $\kB T=1\Gamma$ for \LPT{} (left) and $\kB T=0.2 \Gamma$ for TraSPI (centre). 
    The black solid lines mark the resonances of dot $\da$ (vertical lines) and dot $\db$ (horizontal lines) for particle-like ($\epsilon_q=0$) and hole-like ($\epsilon_q=-U_q$) excitation. 
    The right panel shows the 
    energy scheme for the depicted configurations at $\epsilon_\da=-U_\da/2$ and $\epsilon_\db=0$ (black stars).
    Fermi-distributed occupied lead states are shown in orange for \LPT{} and as black dashed line for TraSPI, the dot levels are shown as thick black lines.}
\end{figure*}
To validate the \LPT{} results for finite Coulomb interaction, we can compare to the numerically-exact method TraSPI \cite{mundinar_transfer-matrix_2022}, which is able to take all orders of coupling strength into account for one interacting and one non-interacting QD (here $U_\da\neq0$ and $U_\db=0$).
Figs.~\ref{fig:TraSPI} (left) and (centre) present the conductance for $\phi=\pi$ and Coulomb interactions $U_\da=4.5\Gamma$ and $U_\db=0$.
The left panel shows results from \LPT{} at a thermal energy of $\kB T=\Gamma$, which is just at the border where linear perturbation theory is applicable, while the centre panel displays existing data obtained using the TraSPI method at a lower temperature of $\kB T=0.2 \Gamma$, such that higher-order processes become important.
Qualitatively, both methods yield as the same key feature conductance peaks near the resonances $\epsilon_\da=\epsilon_\db=0$ and $\epsilon_\da+U_\da=\epsilon_\db=0$, i.e., the crossing points of the black lines.
The broader peak widths observed in the \LPT{} calculations arise from the higher temperature.

Linear perturbation theory predicts \textit{two} conductance valleys due to destructive interference exactly along the diagonals $\epsilon_\da=\epsilon_\db$ and $\epsilon_\da+U_\da=\epsilon_\db$.
The particle-hole symmetric point $2\epsilon_\da+U_\da=0$ and $\epsilon_\db=0$ lies in between these valleys and shows an enhanced conductance.
The energy scheme of this point marked by the black star is depicted in fig.~\ref{fig:TraSPI} (right) and illustrates the lack of destructive interference due to the distance between the energy levels.

The conductance behaviour is different for the TraSPI results.
There, only \textit{one} valley due to destructive interference is visible.
It goes through the particle-hole symmetric point and has a reduced slope as compared to the valleys obtained with \LPT{}.
The difference is due to higher-order processes that are included in TraSPI, with the effect that destructive interference can already occur once the resonance conditions $\epsilon_\da=\epsilon_\db$ and $\epsilon_\da+U_\da=\epsilon_\db$ are already fulfilled within the level broadening set by $\Gamma$.
Then, in fig.~\ref{fig:TraSPI} (right) the narrow energy levels become broader enabling destructive interference between the distant energy levels.

\section{Conclusions} \label{sec:Conclusion}
We have developed a complete theory of electron transport through a double-dot Aharonov-Bohm interferometer up to linear order in the tunnelling-coupling strength $\Gamma$, based on a diagrammatic real-time technique for which we automate the evaluation of the necessary diagrams.
We derive remarkably simple expressions for the dots' occupation and the current, where the phase dependence is completely described by a single complex quantity denoted \textit{effective coherence}.

The theory allows for the variation of all system parameters without assuming any symmetries of the setup.
In particular, finite values of the charging energies are included in the description.
This gives rise to four instead of only one resonance conditions between the particle-like and hole-like transport channels of the two quantum dots.
We find that destructive interference leads to sharp ridges in the interferometric visibility as a function of the dot-level energies.
The width of this feature scales with $\Gamma$, which can be much smaller than the width $\kB T$ of the conductance peaks.
This rich structure is missed in earlier studies for symmetric choices of the system parameters and extreme limits for the charging energy.

To support our interpretation, we compared our results with exact calculations for the non-interacting case as well as with the numerically-exact TraSPI method, which are both non-perturbative in $\Gamma$.  
The predicted sharp features in the visibility may serve as striking indicators for destructive interference in future experimental studies.

%\enlargethispage{3ex}

\section{Funding}
Financial funding by the Deutsche Forschungsgemein\-schaft (DFG, German Research Foundation) under Projects No.\ HU~2303/2-1 and 278162697 - SFB 1242 is acknowledged.

\section{Data availability}
The data underlying this article can be reproduced by the notebook available in the TUDOdata repository, at (link will be added in the editing process).

\section{Conflicts of interest}
The authors declare that they have no competing interests.

\bibliographystyle{oup-unsrt}
\bibliography{ABPaper-editable}

\end{document}